# Anomalous thermal Hall effect in the topological antiferromagnetic state

Kaori Sugii,[1,*] Yusuke Imai,[1] Masaaki Shimozawa,[1,*] Muhammad Ikhlas,[1] Naoki Kiyohara,[1] Takahiro Tomita,[1] Michi-To Suzuki,[2,†] Takashi Koretsune,[3] Ryotaro Arita,[2] Satoru Nakatsuji,[1,4] and Minoru Yamashita[1]

[1]*The Institute for Solid State Physics, The University of Tokyo, Kashiwa, Chiba 277-8581, Japan.*

[2]*RIKEN center for Emergent Matter Science, Wako, Saitama 351-0198, Japan.*

[3]*Department of Physics, Tohoku University, Sendai, 980-8578, Japan.*

[4]*CREST, Japan Science and Technology Agency, Kawaguchi, Saitama 332-0012, Japan.*

[†]*Present address: Institute for Materials Research, Tohoku University, Sendai, Miyagi 980-8577, Japan.*

[*]Corresponding authors.

Email: sugii@issp.u-tokyo.ac.jp (K.S.) and shimo@issp.u-tokyo.ac.jp (M.Shimozawa)

**The anomalous Hall effect (AHE), a Hall signal occurring without an external magnetic field, is one of the most significant phenomena. However, understanding the AHE mechanism has been challenging and largely restricted to ferromagnetic metals. Here, we investigate the recently discovered AHE in the chiral antiferromagnet $Mn_3Sn$ by measuring a thermal analog of the AHE, known as an anomalous *thermal* Hall effect (ATHE). The amplitude of the ATHE scales with the anomalous Hall conductivity of $Mn_3Sn$ over a wide temperature range, demonstrating that the AHE of $Mn_3Sn$ arises from a dissipationless intrinsic mechanism associated with the Berry curvature. Moreover, we find that the dissipationless AHE is significantly stabilized by shifting the Fermi level toward the magnetic Weyl points. Thus, in $Mn_3Sn$, the Berry curvature emerging from the proposed magnetic Weyl fermion state is a key factor for the observed AHE and ATHE.**

## INTRODUCTION

An applied magnetic field perpendicular to the current deflects charge carriers via the Lorentz force, giving rise to a Hall voltage that is linearly proportional to the field. This is the mechanism of the so-called ordinary Hall effect. In contrast, the anomalous Hall effect (AHE) emerges from the following two mechanisms (*1*): one is the intrinsic AHE due to the Berry curvature acting as a fictitious field to Bloch electrons (*2, 3*), and the other is the extrinsic AHE caused by either a skew scattering (an asymmetric impurity scattering) (*4*) or a side-jump mechanism (a sudden shift in the electrons' coordinates during impurity scattering) (*5*) via spin–orbit coupling. According to previous theoretical and experimental studies of various ferromagnetic metals (*6–14*), any mechanism usually generates an anomalous Hall resistivity that is proportional to its magnetization, leading to the prospect that no AHE occurs in antiferromagnetic metals that have zero net magnetization.

Recently, $Mn_3Sn$ has been reported as the first case of an antiferromagnet that exhibits the AHE (*15–17*). $Mn_3Sn$ is a hexagonal antiferromagnet with a stacked kagomé lattice of Mn atoms. Below the Néel temperature of 430 K, Mn magnetic moments of ~3 $\mu_B$ lie in the *ab* plane and form an inverse triangular structure with a negative vector chirality, resulting in a noncollinear antiferromagnetic ordered state (Fig. 1A) (*18*). In this configuration, a tiny in-plane ferromagnetic moment of ~0.002 $\mu_B$/Mn is induced because of spin canting away from the inverse triangular structure; its direction can be controlled by rotating a magnetic field in the *ab* plane (for instance, see Fig. 1A). Most notably, in the antiferromagnetic state, the anomalous Hall conductivity does not scale linearly with magnetization, and its magnitude is comparable to or larger than those of most ferromagnets despite the negligible small magnetization (*15–17*). This is surprising and requires a unified theoretical framework for understanding the AHE in both conventional ferromagnetic metals and $Mn_3Sn$.

One of the most plausible models is the cluster multipole theory (*19*), in which the cluster multipole moments are defined as order parameters to quantify the symmetry breaking accompanied by noncollinear AFM order. Based on this theory, the noncollinear antiferromagnetic order of $Mn_3Sn$ can be regarded as the cluster octupole order that breaks the magnetic symmetry exactly the same as that of the ferromagnetic order, which induces the dissipationless intrinsic AHE obtained by the integrated Berry curvature over the Fermi sea. In fact, it has been theoretically proposed that in $Mn_3Sn$, the Berry curvature generates a large anomalous Hall conductivity, the value of which is roughly consistent with that from experimental results (*19–22*). Moreover, recent *ab initio* studies, ARPES and magnetoresistance measurements of $Mn_3Sn$ (*16*, *20*, *23*, *24*) have revealed the presence of magnetic Weyl points near $E_F$, which provides the possibility that the dissipative AHE identically vanishes (*25*, *26*). Therefore, the observed antiferromagnetic AHE in $Mn_3Sn$ is most likely to show the intrinsic dissipationless nature. However, it has not been determined whether the AHE of the $Mn_3Sn$ is indeed induced by the dissipationless intrinsic mechanism associated with the Berry curvature, and if so, how the proposed magnetic Weyl fermion state in $Mn_3Sn$ is related to the antiferromagnetic AHE.

Here, by performing both thermal and electrical Hall measurements of $Mn_3Sn$, we have investigated the anomalous Hall Lorenz number, $L_{zx}^{AH} = \frac{\kappa_{zx}^{AH}/T}{\sigma_{zx}^{AH}} \left(\frac{e}{k_B}\right)^2$, (where $\kappa_{zx}^{AH}$ ($\sigma_{zx}^{AH}$) represents the anomalous thermal (electrical) Hall conductivity, $k_B$ is the Boltzmann constant, and $e$ is the elementary electric charge), which is sensitive to the AHE mechanism (*8–10*) (see Fig. 1B for the experimental setup). In contrast to the electric current, the heat current of conduction electrons is dissipated by phonons through inelastic scattering; therefore, the anomalous Hall Lorenz number shows a clear difference between dissipationless intrinsic mechanism and dissipative mechanism around one-sixth of the Debye temperature (*8–10*) (for $Mn_3Sn$, $\Theta_D \sim 300$ K (*18*)). However, a previous study of $Mn_3Sn$ by Li *et al.* (*27*) was limited to measurements above 220 K because their sample underwent an additional transition from a noncollinear antiferromagnetic structure to a helical spin structure at ~200 K (*28*). To clarify the mechanism of the AHE at the noncollinear antiferromagnetic state (the cluster octupole ordered state) more precisely, we have determined the anomalous Hall Lorenz number below ~200 K by using $Mn_3Sn$ without the additional magnetic transition (for sample preparation, see MATERIALS AND METHODS). Moreover, we have studied the relation between the proposed magnetic Weyl fermion states and the AHE using the Mn doping dependence of the anomalous Hall Lorenz number, where an extra Mn atom can change $E_F$ toward the Weyl points (*16*). Of note, Mn$_3$Ge is the second reported case of a large AHE in an antiferromagnetic state (*29*, *30*); however, this compound has been suggested to have mainly two types of sources of the Berry curvature near $E_F$, namely band gaps induced by spin–orbit interaction (*31*) and magnetic Weyl points (*23*); hence, it is difficult to investigate the relation between the proposed magnetic Weyl points and the antiferromagnetic AHE (*31*). In this study, a single crystal, Mn$_{3+x}$Sn$_{1-x}$, with two different components (*x* = 0.06 and 0.09, denoted as samples #1 and #2, respectively) has been used. For convenience, these samples are called Mn$_3$Sn in the paper.

## RESULTS

### Longitudinal and transverse electrical transports

We first examine the effect of the extra Mn on the electrical transport in Mn$_3$Sn. Figure 1C shows the temperature dependence of the longitudinal resistivity, $\rho_{xx}(T)$ and $\rho_{zz}(T)$, in sample #1. Here $\rho_{ij}$ is defined as $(-dV/dj)/J_i$ for $\boldsymbol{H} \parallel [01\bar{1}0]$ (*y* axis), where $J_i$ is the electrical current density flowing along the $i = (x, z)$ axis and $(-dV/dj)$ represents the electric field parallel to the $j = (x, z)$ axis. Both $\rho_{xx}(T)$ and $\rho_{zz}(T)$ are similar to the previous results of Mn$_{3.02}$Sn$_{0.98}$ (*15*) except for the residual resistivity ratio (*RRR*); as more excess Mn is added, the *RRR* becomes lower (see Fig. 1D). This result indicates that the extra Mn scatters the charge carriers.

This scattering essentially does not affect the nature of the AHE for Mn$_3$Sn. In our samples, the field variation of the transverse resistivity, $\rho_{zx}(H)$, is not proportional to that of the magnetization (Fig. 2A). Moreover, the absolute value of the Hall conductivity, $\sigma_{zx}$, at 0 T is very large despite the negligibly small remnant magnetization and is roughly estimated to be 90–140 $\Omega^{-1}$cm$^{-1}$ at approximately 100 K from the relation $\sigma_{zx} \equiv -\rho_{zx}/(\rho_{zz}\rho_{xx})$ (see Fig. 2, B and C). Thus, our samples (#1 and #2) show a similar AHE to the previous sample of Mn$_{3.02}$Sn$_{0.98}$ (*15*), supporting that the AHE of Mn$_3$Sn is basically independent of the scattering rates of the charge carriers, $\hbar/\tau$ ($\tau$ is the electron lifetime, and $\hbar$ is the reduced Planck constant). This feature is consistent with the scattering-free AHE in the moderately dirty regime ($\sigma_{ii} = 1/\rho_{ii} \sim 3 \times 10^3 - 5 \times 10^5\ \Omega^{-1}\text{cm}^{-1}$) (*1, 6, 7, 10*), as discussed later (see also Section S1 and Fig. S1).

**Thermal transport properties**

To establish a background for later discussion of the anomalous Hall Lorenz number, we next investigate the charge carrier contribution ($\kappa_{zx}^{ch}$) to the transverse thermal conductivity ($\kappa_{zx}$) of Mn$_3$Sn. Here, $\kappa_{zx}$ is defined as $-w_{zx}/(w_{zz}w_{xx})$, where $w_{ij} = (-dT/dj)/J_i^Q$ for $\boldsymbol{H} \parallel [01\bar{1}0]$ ($y$ axis), $J_i^Q$ is the heat current density along the $i = (x, z)$ axis, and $(-dT/dj)$ represents the temperature gradient parallel to the $j = (x, z)$ axis. In general, three quasi-particles (phonons, magnetic excitations, and charge carriers) can provide thermal transport of antiferromagnetic metals. In the case of Mn$_3$Sn, however, $\kappa_{zx}$ purely originates from charge carriers ($\kappa_{zx} \sim \kappa_{zx}^{ch}$) because of the following two reasons. First, $d\kappa_{zx}(H)/d(\mu_B H)$ is correlated with $d\sigma_{zx}(H)/d(\mu_B H)$ above ~0.1 T. Second, the value of $\kappa_{zx}(H = 0)$ is very large ($10 \sim 30 \times 10^{-3}\ \text{WK}^{-1}\ \text{m}^{-1}$, see Fig. 2, B and D) compared to the cases of magnons (*32*) and phonons (*33*). Of note, the most recent thermal Hall measurements of insulating polar magnets, (Zn$_x$Fe$_{1-x}$)$_2$Mo$_3$O$_8$ (*34*), have shown that the ATHE of magnetic excitations is comparable to our results; however, magnetic excitations of Mn$_3$Sn would not contribute to $\kappa_{zx}$ because they do not carry longitudinal heat current (for details, see Section S2 and Fig. S2A). Therefore, we can consider $\kappa_{zx} \sim \kappa_{zx}^{ch}$ for Mn$_3$Sn, which is significantly different from the longitudinal thermal conductivity of Mn$_3$Sn where phonons are dominant (Section S2 and Fig. S2).

**Anomalous Hall Lorenz number**

To clarify the mechanism of AHE in Mn$_3$Sn, we now compare two anomalous Hall components that are associated with charge carriers, $\sigma_{zx}^{AH}(T)$ and $\kappa_{zx}^{AH}(T)/T$, which are defined as the values of $\sigma_{zx}$ and $\kappa_{zx}^{ch}/T$ ($\sim \kappa_{zx}/T$ for $Mn_3Sn$, as demonstrated above) at $0\,T$ after applying a magnetic field of $0.5\,T$, respectively. In both samples #1 and #2, $\sigma_{zx}^{AH}(T)$ increases monotonically with decreasing temperature, followed by a hump at ~50 K (Fig. 3, A and B); this temperature dependence is qualitatively identical to that of previous reports (*16*), although the amplitude of $\sigma_{zx}^{AH}(T)$ remains ambiguous because of the inaccuracy in the sample dimensions. Similarly, $\kappa_{zx}^{AH}(T)/T$ is enhanced with decreasing temperature and achieves a maximum value at approximately 50 K (Fig. 3, A and B). The observed peak structure reflects the magnetic transition to a cluster-glass phase in which Mn spins are slightly canted toward the *c* axis (*35*). Interestingly, the $\sigma_{zx}^{AH}(T)$ well scales with $\kappa_{zx}^{AH}(T)/T$ for sample #2, whereas it does not scale considerably for sample #1. To illustrate this point, we focus on the anomalous Hall Lorenz number, which is defined as $L_{ij}^{AH} = \frac{\kappa_{ij}^{AH}/T}{\sigma_{ij}^{AH}} \left(\frac{e}{k_B}\right)^2$.

Figure 4A shows the temperature dependence of $L_{ij}^{AH}(T)$ for samples #1 and #2. We also include the two extreme cases, $L_{ij}^{AH}(T)$ for Fe, in which an extrinsic AHE (skew scattering) occurs because of the high conductivity below ~100 K (*9*, *10*), and $L_{ij}^{AH}(T)$ for Ni, which has a dissipationless intrinsic AHE (*8*, *10*). For sample #1, $L_{zx}^{AH}$ becomes $L_0$ below ~$0.2\theta_D$, although it slightly decreases with increasing temperature, whereas for sample #2, $L_{zx}^{AH}$ is virtually temperature independent ($\sim L_0$) up to $0.5\theta_D$. In both cases, $L_{zx}^{AH}$ shows a clearly different temperature dependence from $L_{xy}^{AH}$ for iron (Fe); instead, it is similar to the results of nickel (Ni). This result demonstrates that an inelastic scattering contributes little to the AHE of $Mn_3Sn$ over a wide temperature range (for details, see MATERIALS AND METHODS), indicating that the AHE of $Mn_3Sn$ has an intrinsic dissipationless origin that is associated with the Berry curvature. Here, it should be noted that we can safely exclude the possibility of a side-jump AHE caused by spin−orbit interactions in $Mn_3Sn$ because the interaction of Mn $3d$ electrons is very weak compared to that of Pd $4d$ electrons in the $L1_0$-ordered FePd where the side-jump contribution is dominant (Section S3).

## DISCUSSION

A key question is how the dissipationless intrinsic AHE is related with the proposed magnetic Weyl fermion states in the antiferromagnet $Mn_3Sn$. Previous theoretical studies on conventional ferromagnetic metals (*1, 6, 7*) have pointed out that both the mechanism and magnitude of the AHE are very closely related to the longitudinal electrical conductivity, $\sigma_{ii}$ (Section S1 and Fig. S1). For example, in the moderately dirty regime ($\sigma_{ii} \sim 3 \times 10^3 - 5 \times 10^5 \ \Omega^{-1} cm^{-1}$), the intrinsic AHE is dominant, and its magnitude becomes $\sigma_{ij}^{AH} \approx 6 \times 10^1 - 1 \times 10^3 \ \Omega^{-1} cm^{-1}$. These relations are satisfied in our samples; therefore, at first glance, the AHE of $Mn_3Sn$ seemed to be described within the framework of ferromagnetic metals. However, the impurity effects of $L_{ij}^{AH}$ for $Mn_3Sn$ are drastically different compared with those of ferromagnetic metals.

In ferromagnetic metals with an intrinsic AHE, the increased $\hbar/\tau$ suppresses the value of $L_{xy}^{AH} \sim L_0$ at intermediate temperatures ($0 < T < \theta_D$) and eventually breaks the Wiedemann–Franz law (for example, see the inset of Fig. 4A) (*10*). This indicates that inelastic scattering due to impurities converts the non-dissipative AHE into a dissipative AHE. This experimental result is explained by the following model (*10*): a spin–orbit interaction opens the energy gap, $\Delta^{gap}$, at $E_F$, near which a Berry curvature becomes finite; as a result, a purely intrinsic AHE is possible only for a small inelastic scattering rate ($\hbar/\tau < \Delta^{gap}$); however, in the case of $\hbar/\tau \geq \Delta^{gap}$, the process of inelastic scattering is relatively advanced, resulting in the crossover from the scattering-free AHE to a scattering-dependent one (for details, see Section S4 and Fig. S3, A to C). We now apply this model to the present system. In $Mn_3Sn$, $\hbar/\tau$ is roughly estimated to be $\hbar/\tau^{Drude} \sim 0.25$–$3.1$ eV ($0.35$–$0.88$ eV) based on the two-band (single-band) Drude model by assuming that $m$ is equal to the mass of the free electron (Section S5). The value of $\Delta^{gap}$ for $Mn_3Sn$, if it exists, is expected to be the same order of magnitude as that of ferromagnetic metals (for example, $\Delta^{gap} \sim 70$ meV for Fe (*13*)) because the spin–orbit interaction of $3d$ ions is comparable with one another. Therefore, we expected a large suppression of $L_{zx}^{AH}$ in $Mn_3Sn$ because of $\hbar/\tau \gg \Delta^{gap}$. Nevertheless, Mn$_3$Sn exhibits the opposite trend (Fig. 4A): $L_{zx}^{AH}$ is approximately close to $L_0$ in the wide temperature range, although $\sigma_{ii}$ is one or two orders of magnitudes lower than the case of Ni; more surprisingly, $L_{zx}^{AH}$ gets closer to $L_0$ by the extra Mn that acts as impurity scattering for Mn$_3$Sn. These results point to the possibility that another scenario beyond the Drude model explains the dissipationless intrinsic AHE of Mn$_3$Sn.

According to our first-principles calculations, Mn$_3$Sn harbors a magnetic Weyl fermion state that has the energy bands with Weyl points near $E_F$ (Weyl bands) and the other metallic bands (see Fig. 4, B and C). This is consistent with previous reports of other groups (*20, 23*). Theoretically, the scattering rate of the Weyl band, $\hbar/\tau^{Weyl}$, is significantly different from $\hbar/\tau$ calculated via the Boltzmann transport theory (*36*), thus supporting the failure of the Drude model in Mn$_3$Sn. In addition, it has recently been suggested that the intrinsic AHE is induced by the Weyl bands and that the dissipative AHE completely vanishes when $\hbar/\tau$ is smaller than the cutoff energy, $\Delta^{cut}$, which is defined as the energy range of a linear dispersion (see Section S4 and Fig. S3, D to F) (*25, 26*). Therefore, the observed dissipationless intrinsic AHE in Mn$_3$Sn shows the presence of Weyl bands with $\hbar/\tau^{Weyl} < \Delta^{cut}$ ($\hbar/\tau^{Weyl}$ is roughly estimated to be 14.4–21.4 meV; Section S6). In contrast to $\hbar/\tau^{Weyl}$, the scattering rate of the charge carriers in the metallic bands is expected to be nearly equal to $\hbar/\tau^{Drude}$ because the Weyl bands have little effect on the longitudinal electrical transport because of its small carrier density. Indeed, it has been reported that in Mn$_3$Sn, a negative magnetoresistance associated with a chiral anomaly is very small compared to the case of other Weyl semimetals without metallic bands (*24*). Thus, a magnetic Weyl metal with two clearly different scattering times ($\tau^{Weyl}$ and $\tau^{Drude}$; see Fig. 4D) is fully consistent with our experimental results.

Our results indicate that in Mn$_3$Sn, a finite Berry curvature that emerges from the magnetic Weyl fermion state is a key source for the observed large AHE and ATHE. This strongly supports the previous conclusion that the large anomalous Nernst effect (ANE) of Mn$_3$Sn is associated with a Berry curvature (*16*). Moreover, we found that excess Mn in Mn$_3$Sn acts as a scattering of thermal transport, even though it increases the signal of the ANE (*16*). This is useful for the realization of a thermoelectric device based on ANE. Further experimental and theoretical studies are required for understanding magnetic Weyl metals, such as Mn₃Sn, to develop functional devices including thermoelectric devices.

## MATERIALS AND METHODS

### Sample growth

Two single crystals of $Mn_{3+x}Sn_{1-x}$ ($x = 0.06$ and $0.09$) were synthesized by the Bridgman method described in Ref. (*16*). This method avoids an additional transition at $T \sim 200$ K (*15*, *16*). The magnetization measurements were performed using a commercial SQUID magnetometer (MPMS, Quantum Design) and confirmed the absence of the additional transition in the quenched samples.

### Electrical and thermal transport measurements

The electrical transport properties were measured using the standard four-probe method (see Fig. 1B for the setup for the longitudinal ($\Delta V_L$) and transverse ($\Delta V_T$) voltage probes). Electrical contacts were made by spot-welding gold wires to reduce the contact resistance. The thermal transport measurements were performed via the standard steady-state method. We attached three Cernox thermometers (CX1050) and one heater onto the samples through gold wires (see Fig. 1B for the configuration of the three thermometers: $T_{High}$, $T_{L1}$, and $T_{L2}$). Thermal contacts between the samples and thermometers were made by using the same gold wires with electrical contacts. To avoid both electrical and thermal Hall signals from the metals used in the cryostat, we used an insulating LiF single crystal as a heat bath. In addition, a nonmagnetic silicon grease was used to attach the sample to the LiF heat bath (*33*, *37*). The electrical (heat) current was applied along the $[2\bar{1}\bar{1}0]$ ($x$ axis) or $[0001]$ ($z$ axis) axis. The magnetic field was applied along the $[01\bar{1}0]$ ($y$ axis) axis and swept between $+0.5$ and $-0.5$ T at a sweep rate of $0.0167$ T/min. The electric Hall conductivity, $\sigma_{zx}$, and thermal Hall conductivity, $\kappa_{zx}$, were obtained from the following relations,

$$\sigma_{zx} \equiv -\rho_{zx}/(\rho_{zz}\rho_{xx}) = \sigma_{xx} \ (\Delta V_T/w)/(\Delta V_L/l) \qquad \text{and} \qquad \kappa_{zx} \equiv -w_{zx}/(w_{zz}w_{xx}) =$$

$\kappa_{xx}$ $(\Delta T_T/w)/(\Delta T_L/l)$, where $\Delta T_T = T_{L1} - T_{L2}$, $\Delta T_L = T_{High} - T_{L1}$, $\sigma_{xx} = -J_x/(\Delta V_L/l)$, and $\kappa_{xx} = -J_x^Q/(\Delta T_L/l)$; moreover, $w$ is the sample width, $l$ is a distance between thermal contacts of the two thermometers ($T_{High}$ and $T_{L1}$), and $J_x$ ($J_x^Q$) is the electrical (heat) current density flowing along the $x$ axis. To take into account the anisotropy between $\rho_{xx}$ and $\rho_{zz}$ ($\kappa_{xx}$ and $\kappa_{zz}$), we measured the longitudinal electrical (thermal) conductivities in both the $x$ and $z$ directions in the same sample.

**Anomalous Lorenz Hall number**

For intermediate temperatures ($0 < T < \theta_D$), the value of $L_{ij}^{AH}$ is strongly affected by the mechanism of the AHE. The Wiedemann–Franz law applies ($L_{ij}^{AH} \sim L_0$) when the dissipationless intrinsic AHE is dominant (*8, 10*). In contrast, a large reduction in $L_{ij}^{AH}$ has been reported for the extrinsic AHE because of skew scatterings (inelastic scatterings) (*9, 10*); in particular, a clear change of $L_{ij}^{AH}$ is expected to appear at approximately $T \sim \theta_D/6$ (for example, see the result for Fe in Fig. 4A). In this work, to determine the dominant contribution to the AHE in Mn$_3$Sn, we measured the temperature dependence of $L_{ij}^{AH}$ over a wide temperature range that includes $\theta_D/6 \sim 50$ K for Mn$_3$Sn (*18*).

**DFT calculations**

As described in Ref. (*24*), electronic structures were calculated for the noncollinear antiferromagnetic state of Mn$_3$Sn using the QUANTUM ESPRESSO package using a relativistic version of the ultrasoft pseudo potentials. We used the lattice constants obtained via powder X-ray measurements at 60 K (*24*) and the Wyckoff position of the Mn 6h atomic sites of $x = 0.8388$ (*18*). The calculated results were obtained for the magnetic texture for $\boldsymbol{H}$ // $y$ (see Fig. 1A).

## SUPPLEMENTARY MATERIALS

Supplementary material for this article is available at http://xxx.

Section S1. Analysis of $\sigma_{ii}$ versus $\sigma_{ij}^{AH}$

Section S2. Longitudinal thermal transport in the antiferromagnetic metal, Mn$_3$Sn

Section S3. Side-jump contribution to the AHE in Mn$_3$Sn

Section S4. Crossover from non-dissipative to dissipative AHEs

Section S5. Estimation of the scattering time from the Drude model

Section S6. Estimation of the scattering time of Weyl bands

Fig. S1. $\sigma_{ij}^{AH}$ versus $\sigma_{ii}$ for ferromagnetic metals and Mn$_3$Sn.

Fig. S2. Longitudinal thermal transport properties of Mn$_3$Sn.

Fig. S3. Intrinsic AHE in conventional ferromagnetic metals and Weyl semimetals.

Fig. S4. Energy dependence of calculated density of states (DOS) near the Fermi energy ($E_{\mathrm{F}}$).

References (*38–44*)

**Acknowledgements:** We thank Hua Chen, S. Fujimoto, P. Goswami, T. Higo, M. Oshikawa, Y. Shiomi, A. Shitade, Y. Tada, and Y. Yanase for valuable discussions.

**Funding:** This research was supported by Toray Science Foundation, CREST (Grant No. JPMJCR18T3), Japan Science and Technology Agency, PRESTO, Japan Science and Technology Science, Grants-in-Aid for Scientific Research (Grant No. 15K17691, No. 15K17713, No. 16K17743, No. 16H02209, and No. 25707030), Grants-in-Aids for Scientific Research on Innovative Areas "Topological Materials Science" (Grant No. JP18H04213 and No. JP18H04230) and "J-Physics" (Grant No. 15H05882 and No. 15H05883), and Program for Advancing Strategic International Networks to Accelerate the Circulation of Talented Researchers (Grant No. R2604) from the Japanese Society for the Promotion of Science.

**Author contributions:** K.S., M.Shimozawa, S.N., and M.Y. conceived the project. K.S., Y.I., M.Shimozawa, and M.Y. performed the electrical and thermal transport measurements. M.I., N.K.,

T.T., and S.N. carried out sample preparation and characterizations. K.S., M.Shimozawa, and M.Y. contributed to the interpretation of the experimental results based on the first-principles calculation performed by M.Suzuki, T.K., and R.A. K.S., M.Shimozawa, and M.Y. wrote the manuscript with input from S.N.

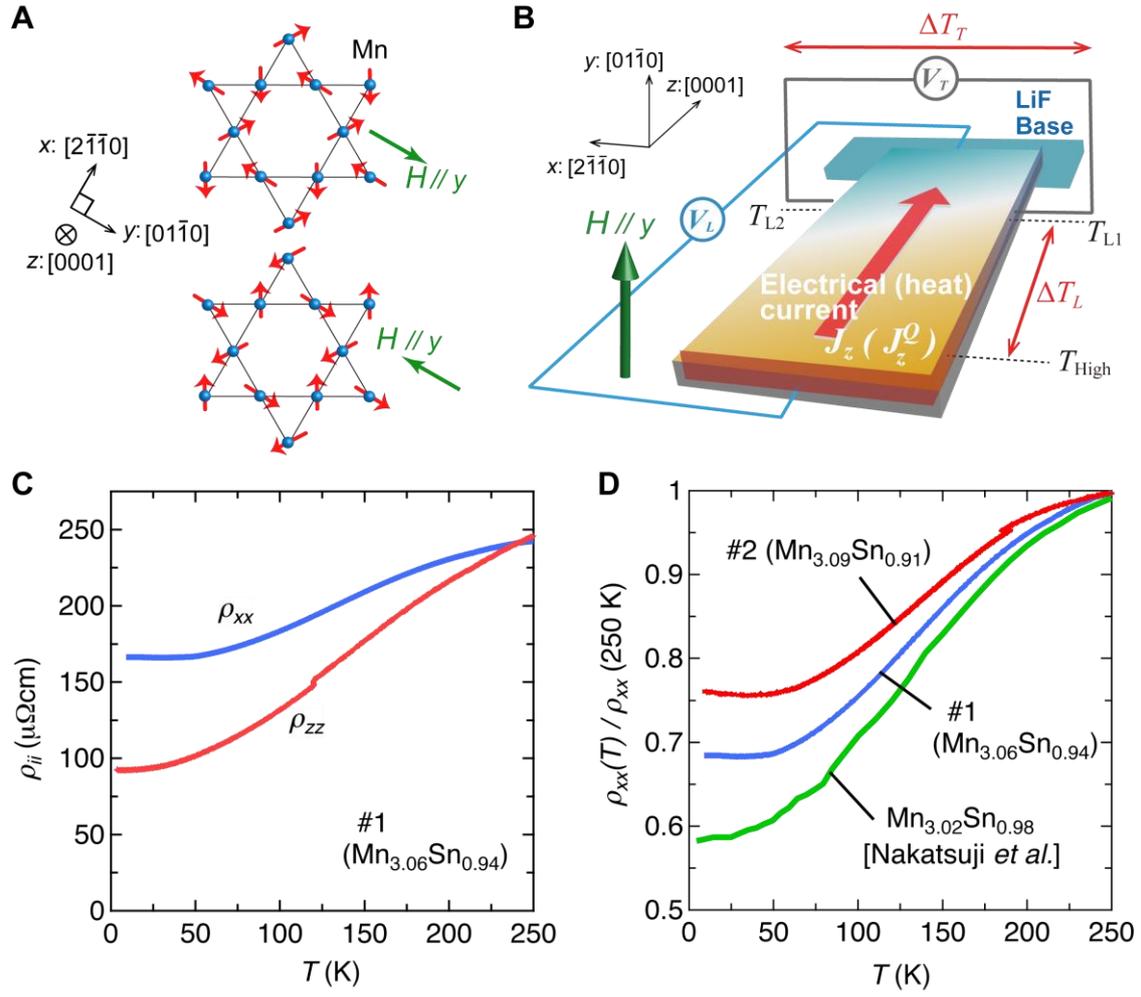

**Fig. 1. Fundamental properties of Mn₃Sn.** (**A**) Schematic Mn spin structures of Mn₃Sn viewed along the [0001] axis. The Mn atoms (blue circles) with spins (red arrows) are located on the kagomé lattice. The upper and lower panels represent the spin configurations for the magnetic field along the $[01\bar{1}0]$ ($//+y$) axis and vice versa (*15*). (**B**) Experimental setup for electrical and thermal Hall measurements (for details, see MATERIALS AND METHODS). (**C**) Temperature dependence of the longitudinal resistivity, $\rho_{xx}(T)$ (blue) and $\rho_{zz}(T)$ (red), in sample #1. (**D**) $\rho_{xx}(T)$ normalized by the value at 250 K, $\rho_{xx}(T)/\rho_{xx}(250\ \text{K})$ for samples #1 (blue) and #2 (red). For comparison, we also plot the same data for Mn₃.₀₂Sn₀.₉₈ (green) (*15*).

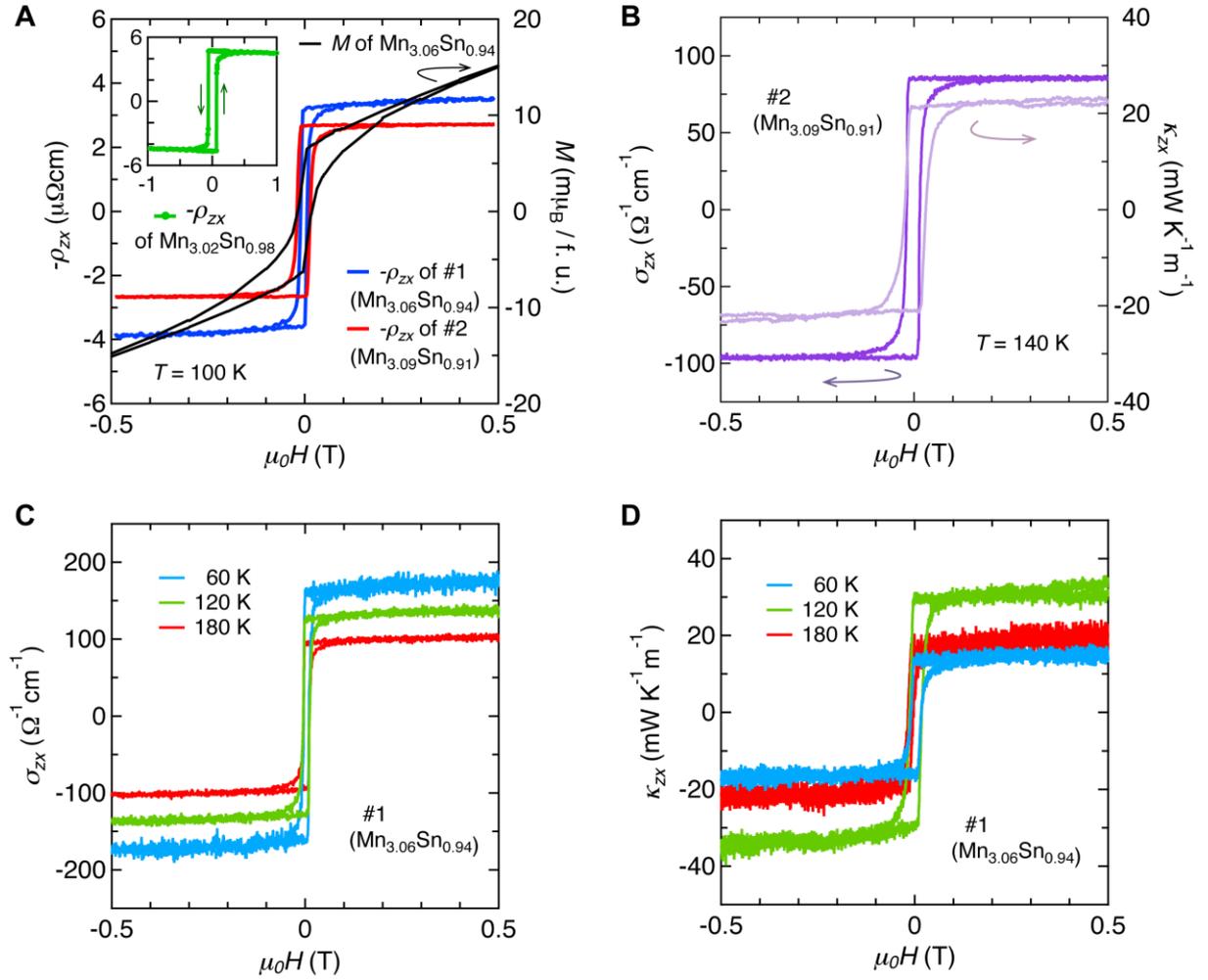

**Fig. 2. Electrical and thermal Hall effects of Mn$_3$Sn.** (**A**) Magnetic field dependence of transverse resistivity, $\rho_{zx}$ (left axis), for samples #1 (blue) and #2 (red) at 100 K. We also plot the magnetic field dependence of magnetization, $M$, at the same temperature for Mn$_{3.06}$Sn$_{0.94}$ taken from Ref. (*16*) (black line, right axis). The inset shows $\rho_{zx}(H)$ for Mn$_{3.02}$Sn$_{0.98}$ taken from Ref. (*15*). (**B**) Magnetic field dependence of the transverse conductivity, $\sigma_{zx}$ (dark purple, left axis), and the transverse thermal conductivity, $\kappa_{zx}$ (light purple, right axis), for sample #2 at 140 K. (**C** and **D**) Temperature variation of $\sigma_{zx}(H)$ (C) and $\kappa_{zx}(H)$ (D) for sample #1.

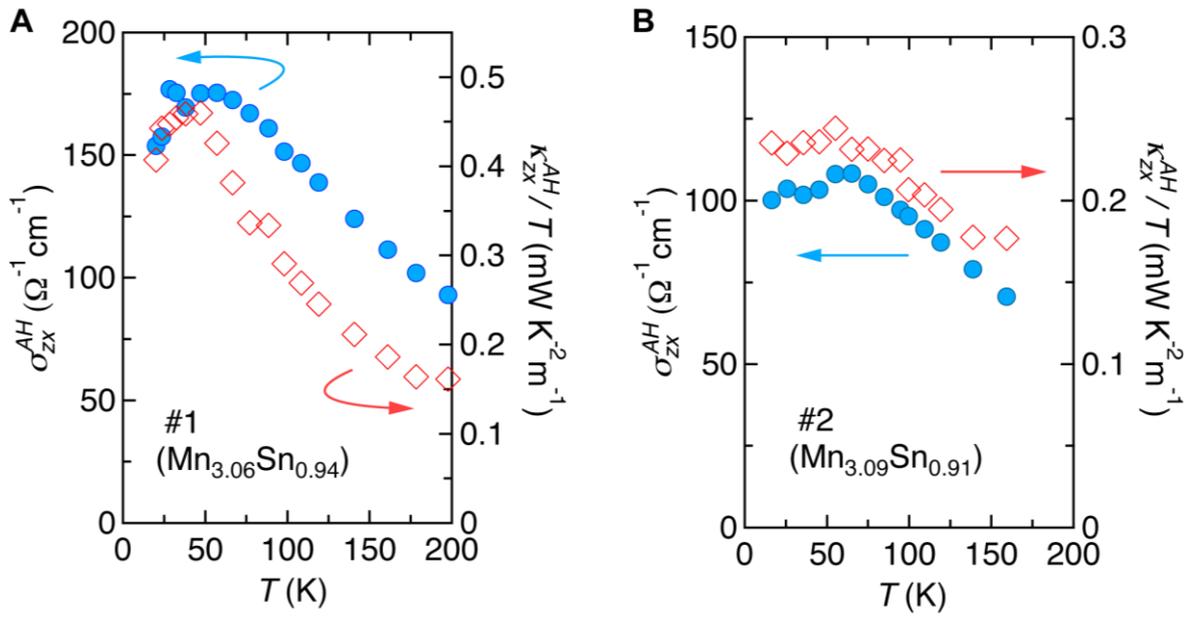

**Fig. 3**. **Anomalous Hall components of Mn₃Sn.** (**A** and **B**) Temperature dependence of the anomalous Hall conductivity, $\sigma_{zx}^{AH}$ (filled circles, left axis), and the anomalous thermal Hall conductivity, $\kappa_{zx}^{AH}/T$ (open diamonds, right axis), for samples #1 (A) and #2 (B).

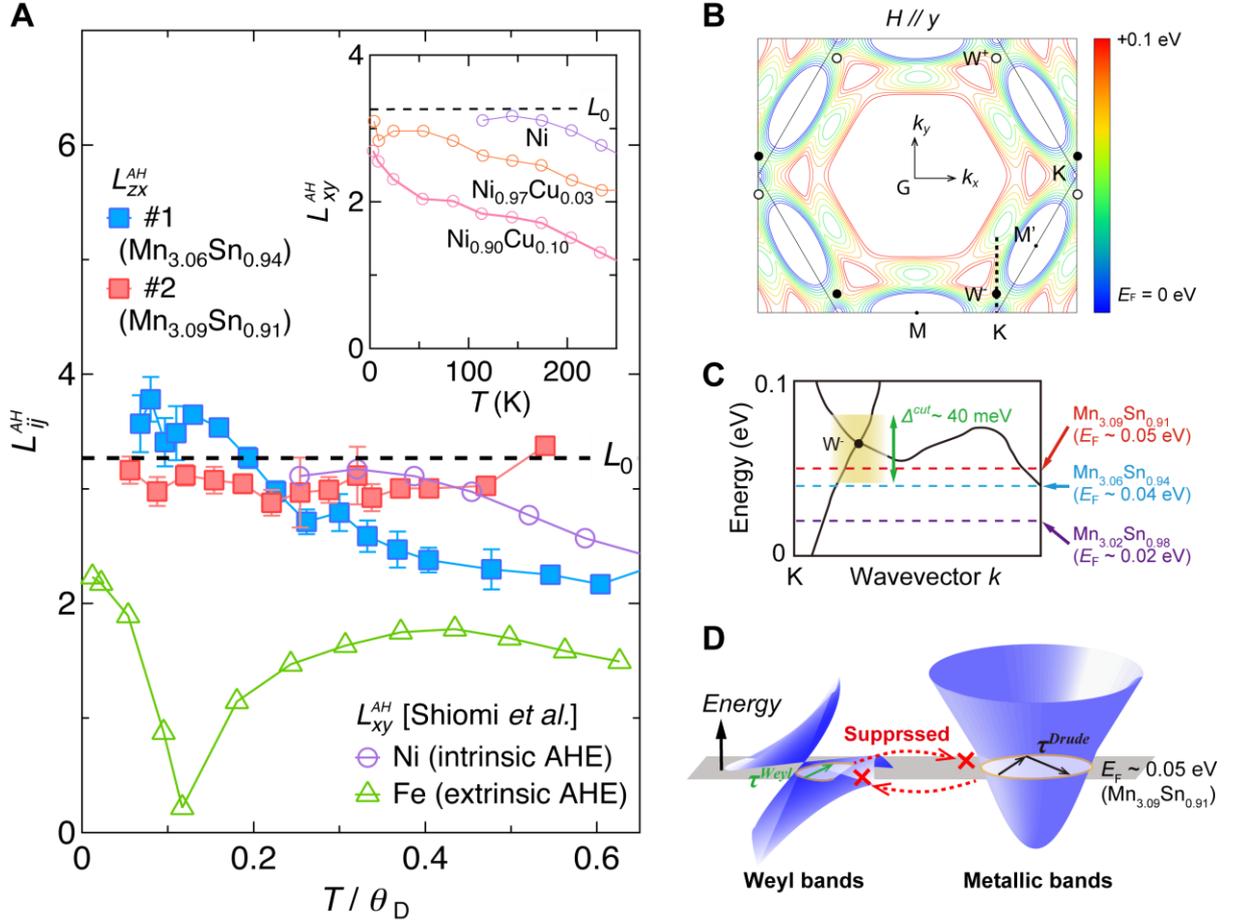

**Fig. 4**. **Anomalous Hall Lorenz number.** (**A**) Temperature dependence of the anomalous Hall Lorenz number, $L_{zx}^{AH}$, for samples #1 (filled blue circles) and #2 (filled red circles). For comparison, we also plot the temperature dependence of $L_{xy}^{AH}$ for the ferromagnetic metals Ni (open purple squares) with $\theta_{\mathrm{D}} \sim 450$ K and fcc Fe (open green triangles) with $\theta_{\mathrm{D}} \sim 470$ K (*9*, *10*). The dashed line represents the value of $L_0 \equiv \pi^2/3$. The inset shows the temperature dependence of $L_{xy}^{AH}$ for Ni, $Ni_{0.97}Cu_{0.03}$, and $Ni_{0.90}Cu_{0.10}$ (*10*). (**B**) Contour plot of the band below the Weyl point for $Mn_{3+x}Sn_{1-x}$ in the $k_x$-$k_y$ plane at $k_z = 0$. The color scale represents $E_{\mathrm{F}}$ for $Mn_{3+x}Sn_{1-x}$ measured from that of the non-Mn doped sample, $Mn_{3.00}Sn_{1.00}$. Each colored line represents the result for the corresponding $E_{\mathrm{F}}$. A pair of two Weyl nodes with different chiralities is shown by the filled and open circles. All of the calculated results were obtained for the magnetic structure shown in Fig. 1A (see also MATERIALS AND METHODS). (**C**) Schematic band dispersion near the Weyl point in $Mn_3Sn$. The horizontal axis represents the wavevector $k$ along the black dashed line shown in (**B**). The purple, blue, and red dashed lines represent the Fermi energies of $Mn_{3.02}Sn_{0.98}$ ($E_{\mathrm{F}} \sim 0.02$ eV), $Mn_{3.06}Sn_{0.94}$ ($E_{\mathrm{F}} \sim 0.04$ eV), and $Mn_{3.09}Sn_{0.91}$ ($E_{\mathrm{F}} \sim 0.05$ eV), respectively (*16*, *24*). The shaded area roughly indicates the energy region of linear dispersion, namely, the Weyl band. (**D**) Schematic band structures of $Mn_{3.09}Sn_{0.91}$. The gray-shaded plane represents $E_{\mathrm{F}}$. The green and black solid (red dashed) lines represent intra-band (inter-band) scatterings. Our study indicates that the scattering rates within the Weyl bands ($\tau^{Weyl}$) and within the metallic bands ($\tau^{Drude}$) are not equal to each other, leading to the conclusion that the inter-band scatterings between Weyl bands and metallic bands are strongly suppressed.

# Supplementary Materials for

# Anomalous thermal Hall effect in the topological antiferromagnetic state

Kaori Sugii,[1,*] Yusuke Imai,[1] Masaaki Shimozawa,[1,*] Muhammad Ikhlas,[1] Naoki Kiyohara,[1] Takahiro Tomita,[1] Michi-To Suzuki,[2,†] Takashi Koretsune,[3] Ryotaro Arita,[2] Satoru Nakatsuji,[1,4] and Minoru Yamashita[1]

[1]*The Institute for Solid State Physics, The University of Tokyo, Kashiwa, Chiba 277-8581, Japan.*

[2]*RIKEN center for Emergent Matter Science, Wako, Saitama 351-0198, Japan.*

[3]*Department of Physics, Tohoku University, Sendai, 980-8578, Japan.*

[4]*CREST, Japan Science and Technology Agency, Kawaguchi, Saitama 332-0012, Japan.*

[†]*Present address: Institute for Materials Research, Tohoku University, Sendai, Miyagi 980-8577, Japan.*

[*]Corresponding authors.

Email: sugii@issp.u-tokyo.ac.jp (K.S.) and shimo@issp.u-tokyo.ac.jp (M.Shimozawa)

## Section S1. Analysis of $\sigma_{ii}$ versus $\sigma_{ij}^{AH}$

According to previous studies of ferromagnetic metals ($6$, $7$), the AHE is classified into three regions on the basis of the magnitude of longitudinal conductivity $\sigma_{ii}$, as shown in Fig. S1. First is the super clean region ($\sigma_{ii} \geq 5 \times 10^6 \ \Omega^{-1}\text{cm}^{-1}$) in which the skew scattering contribution to the AHE is dominant ($\sigma_{ij}^{sk} \sim \sigma_{ij}^{AH}$). The skew scattering depends on the impurity density, leading to $\sigma_{ij}^{AH} \propto \sigma_{ii}$. The second region is known as a moderately dirty metal ($\sigma_{ii} \sim 3 \times 10^3 - 5 \times 10^5 \ \Omega^{-1}\text{cm}^{-1}$). As shown in the main text, the intrinsic AHE that is essentially independent of the impurity density is dominant in this region ($\sigma_{ij}^{AH} \sim \sigma_{ij}^{int} \sim$ const.). The final area is the dirty regime of $\sigma_{ii} < 3 \times 10^3 \ \Omega^{-1}\text{cm}^{-1}$. The AHE of this regime is determined by the intrinsic mechanism, as with the moderately dirty metal; however, $\sigma_{ij}^{AH}$ follows a different relation, specifically $\sigma_{ij}^{AH} \sim \sigma_{ij}^{int} \propto (\sigma_{ii})^{1.6}$, possibly due to dephasing of the Berry phase that is caused by variable range hopping conduction. Thus, a relation between $\sigma_{ij}^{AH}$ and $\sigma_{ii}$ is commonly found in ferromagnetic metals.

We now apply the above discussion to the antiferromagnetic metal, Mn₃Sn. According to our resistivity measurements, $\sigma_{ii}$ of Mn₃Sn is roughly estimated to be $\sim 6 \times 10^3 \, \Omega^{-1}\text{cm}^{-1}$, indicating that Mn₃Sn belongs to the moderately dirty region. In this case, it is expected that the intrinsic AHE is dominant and that its magnitude becomes $\sigma_{ij}^{AH} \approx 6 \times 10^1 - 1 \times 10^3 \, \Omega^{-1}\text{cm}^{-1}$. Surprisingly, these two expectations are fulfilled by our samples despite the antiferromagnetic metal, implying the universal relation between $\sigma_{ii}$ and $\sigma_{ij}^{AH}$ (see Fig. S1).

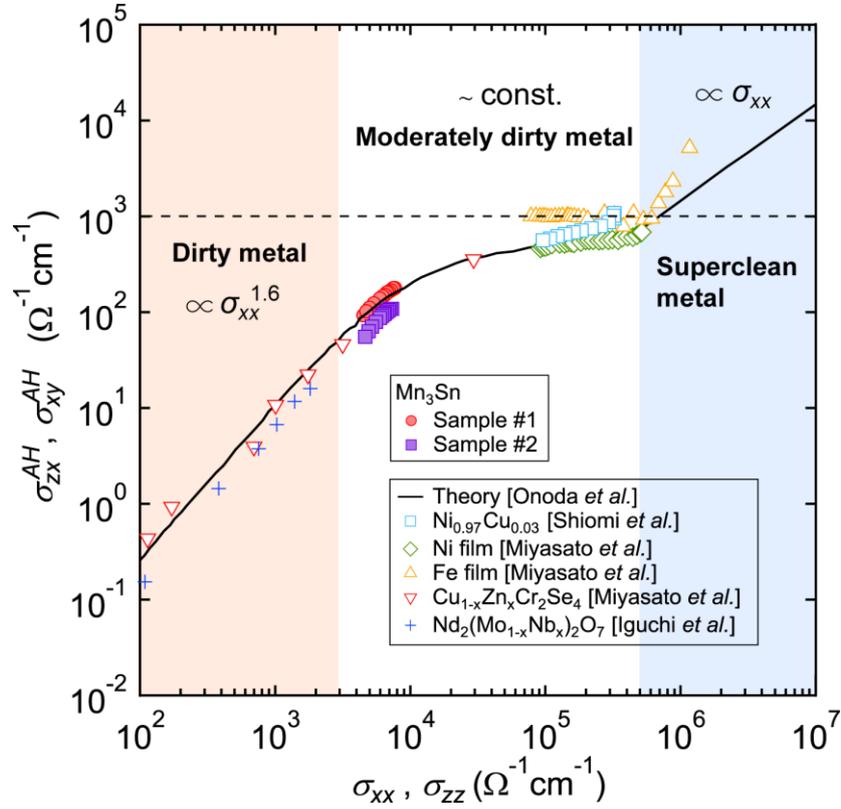

**Fig. S1.** $\sigma_{ij}^{AH}$ **versus** $\sigma_{ii}$ **for ferromagnetic metals and Mn₃Sn.** The filled symbols represent $\sigma_{zx}^{AH}$ versus $\sigma_{zz}$ for Mn₃Sn samples #1 (red circles) and #2 (purple squares). We also present an experimental $\sigma_{xy}^{AH} - \sigma_{xx}$ plot for several ferromagnetic metals (open symbols). These data are taken from Shiomi *et al.* (*10*) for Ni₀.₉₇Cu₀.₀₃ (light blue squares), from Miyasato *et al.* (*38*) for a Ni film (green diamonds), Fe film (yellow triangles), and Cu₁₋ₓZnₓCr₂Se₄ (*x* = 0.0, 0.2, 0.4, 0.5, 0.6, 0.8, and 0.9, red inverted triangles), and from Iguchi *et al.* (*39*) for Nd₂(Mo₁₋ₓNbₓ)₂O₇ (0.01 < *x* ≤ 0.1, blue crosses). The theoretical result calculated by Onoda *et al.* (*7*) is shown as the solid curve. The AHE of ferromagnetic metals are classified into three regimes: dirty metal regime, moderately dirty metal regime, and super clean metal regime.

## Section S2. Longitudinal thermal transport in the antiferromagnetic metal, Mn₃Sn

Generally, longitudinal thermal transport in antiferromagnetic metals provides important information for phonons, magnons, and charge carriers. This is a big advantage compared with the electrical transport but also highlights the difficulty to extract each component from the observed thermal conductivity, $\kappa_{ij}(T)$, which is defined as $-J_i^Q/(dT/dj)$ where $J_i^Q$ is the heat current density along the $i = (x, z)$ axis, and $(-dT/dj)$ represents the temperature gradient parallel to the $j = (x, z)$ axis. However, in Mn₃Sn, the magnon contribution is negligibly small, especially above ~50 K, because $\kappa_{ii}(T)$ is essentially field independent up to 15 T (Fig. S2A). In addition, the possible thermal conductivity of the charge carriers is less than ~35% of the total one above ~20 K even though we assume that the Wiedemann–Franz law holds in this temperature range (Fig. S2B). Therefore, phonons mainly contribute to the $\kappa_{ii}(T)$ of Mn₃Sn above 20 K.

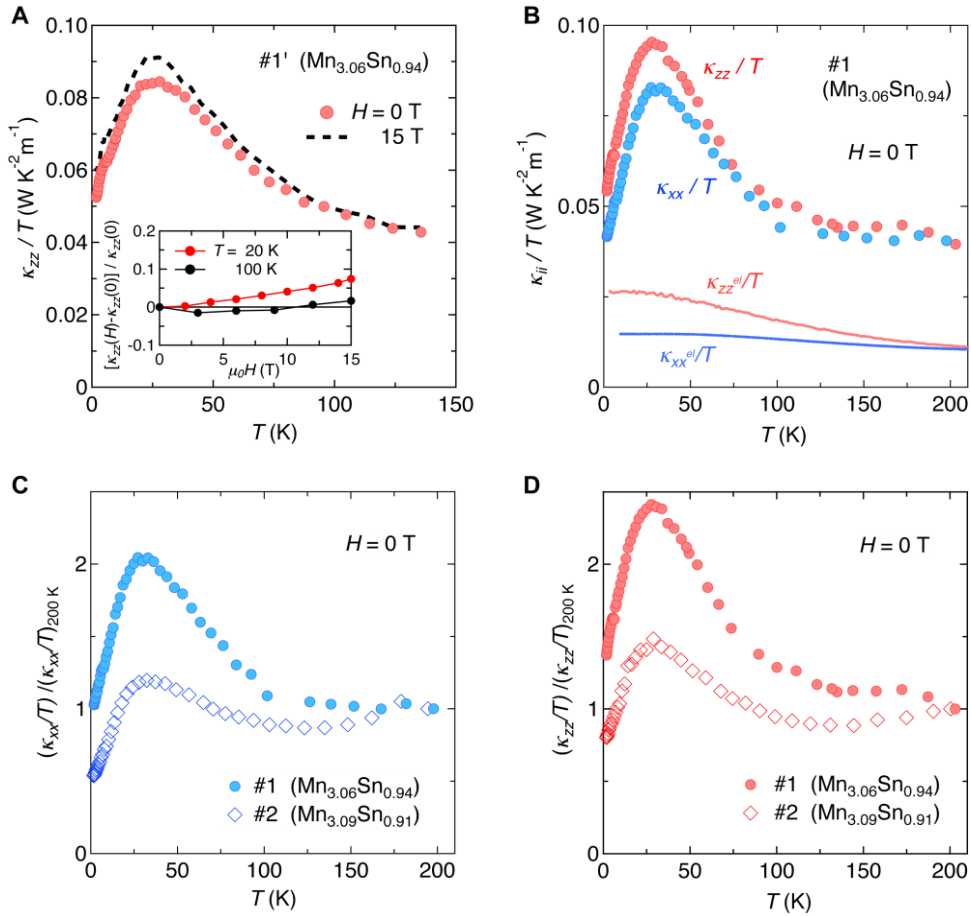

**Fig. S2. Longitudinal thermal transport properties of Mn₃Sn.** (**A**) Temperature dependence of the longitudinal thermal conductivity divided by temperature along the $z$ axis, $\kappa_{zz}(T)/T$, for another single crystal, Mn₃.₀₆Sn₀.₉₄ (sample #1′), which is from the same batch as sample #1. The filled circles and dotted lines represent $\kappa_{zz}(T)/T$ at 0 and 15 T, respectively. The inset plots the magnetic field dependence of the thermal conductivity, $\kappa_{zz}(H)$, normalized by the zero field value $[\kappa_{zz}(H) - \kappa_{zz}(0)]/\kappa_{zz}(0)$ for sample #1′. (**B**) $\kappa_{xx}(T)/T$ (blue circles) and $\kappa_{zz}(T)/T$ (red circles) for sample #1. The blue and red curves represent the contribution of charge carriers to $\kappa_{xx}(T)/T$ and $\kappa_{zz}(T)/T$ calculated from the Wiedemann–Franz law. (**C** and **D**) $\kappa_{xx}(T)/T$ (C) and $\kappa_{zz}(T)/T$ (D) normalized by the value at 200 K for samples #1 (filled circles) and #2 (open diamonds).

Next, we examined the Mn doping dependence of the phonon thermal conductivity for Mn₃Sn. As shown in Fig. S2, a peak for $\kappa_{ii}(T)/T$ clearly appears at approximately 30 K. This peak structure could correspond to a phonon peak, the amplitude of which benchmarks the sample quality (*40*). In sample #2, the phonon peak becomes smaller than that of sample #1 (Fig. S2, C and D). This indicates that excess Mn acts as a scatterer of phonons as well as charge carriers (see also the main text and Fig. 1D).

## Section S3. Side-jump contribution to the AHE in Mn₃Sn

The anomalous Hall conductivity ($\sigma_{ij}^{AH}$) is generally described by the sum of three components: side jump ($\sigma_{ij}^{sj}$), skew scattering ($\sigma_{ij}^{skew}$), and intrinsic ($\sigma_{ij}^{int}$) contributions. Among them, $\sigma_{ij}^{sj}$ and $\sigma_{ij}^{int}$ are independent of the impurity density, making it difficult to separate them. Until now, the side-jump AHE has often been estimated via two different methods. The first involves utilizing the difference between the experimentally observed $\sigma_{ij}^{AH}$ and the theoretically calculated $\sigma_{ij}^{int}$ in the limit of $\sigma_{ii} \to 0$, namely, $\left(\sigma_{ij}^{AH} - \sigma_{ij}^{sj}\right)|_{\sigma_{ii} \to 0} \sim \sigma_{ij}^{int}$. For example, in various 3*d*-ferromagnetic metals and semiconductors (*11–14*), the calculated $\sigma_{ij}^{int}$ agrees well with the observed $\sigma_{ij}^{AH}$, which indicates the dominance of the intrinsic AHE over the side-jump AHE. Another example is the $L1_0$-ordered ferromagnet FePd (*41, 42*); this compound shows a larger scattering-independent $\sigma_{ij}^{AH}$ compared with $\sigma_{ij}^{int}$ estimated via *ab initio* calculations (*14*), suggesting a dominance of the side-jump contribution in the $L1_0$-ordered FePd. The other possible way is via the magnitude of the spin–orbit interaction, which strongly depends on the side-jump AHE. In fact, Pd $4d$ electrons have a relatively large spin–orbit interaction compared with $3d$-ferromagnetic metals. Here, we consider the case of $Mn_3Sn$ from these points of view. As with Mn alloys with a negligibly small side-jump contribution (*11, 43*), $Mn_3Sn$ has a relatively small spin–orbit interaction of Mn $3d$ electrons. Moreover, the value of $\sigma_{ij}^{AH}$ estimated from the *ab initio* calculations (*19–23*) is comparable to the experimental results. Thus, we can rule out the possibility of a side-jump AHE caused by spin–orbit interactions in $Mn_3Sn$.

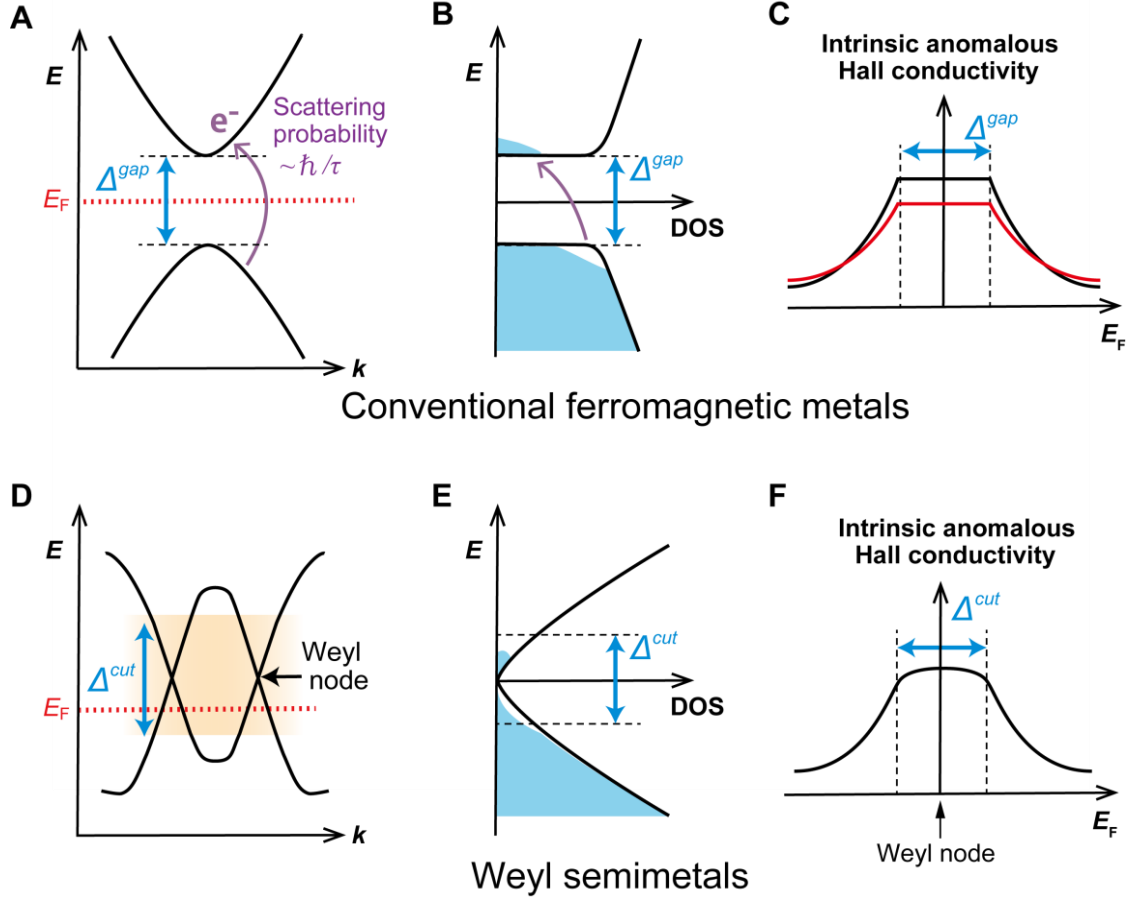

**Fig. S3. Intrinsic AHE in conventional ferromagnetic metals and Weyl semimetals.** (**A** to **C**) Schematic illustrations of band dispersion (A), electron energy distribution (the product of the probability distribution and the density of states (DOS)) (B), and intrinsic anomalous Hall conductivity (C) in a simple model of ferromagnetic metals. The blue shaded region in (B) represents the filled electronic states. In the vicinity of $E_F$, a narrow gap, $\Delta^{gap}$, is opened via spin–orbit coupling. When the relation of $\Delta^{gap} < \hbar/\tau$ is satisfied, electrons are easily scattered from the lower to upper bands, as shown in purple arrows. In this case, the energy dependence of the intrinsic anomalous Hall conductivity shown in (C) is modified from the black line to the red line. (**D** to **F**) Schematic illustrations of band dispersion (D), electron energy distribution (E), and intrinsic anomalous Hall conductivity (F) in a toy model of magnetic Weyl semimetals (*44*). For convenience of discussion, one pair of Weyl points is located in the vicinity of $E_F$. $\Delta^{cut}$ represents the cutoff energy, which is defined as the energy range of linear dispersion (orange shaded area). The blue-shaded region in (E) represents filled electronic states. Interestingly, the energy dependence of the intrinsic anomalous Hall conductivity in (F) (black line) looks very similar to that in (C).

### Section S4. Crossover from non-dissipative to dissipative AHEs

Here, we explain how to interpret the crossover from non-dissipative to dissipative AHEs along with a previous report (*10*). Figure S3A shows the schematic band dispersion of ferromagnetic metals that exhibit the intrinsic AHE associated with a narrow gap that is opened by spin–orbit coupling. For simplicity of discussion, we consider that $E_F$ is located within the gap and that the charge carrier is an electron. According to the uncertainty principle, the decrease in the scattering time ($\tau$) due to an impurity brings about energy broadening of the electronic states ($\Delta E \sim \hbar/\tau$). When the band broadening is smaller than the gap size ($\hbar/\tau < \Delta^{gap}$), the electrons are barely scattered from the lower occupied band to the upper unoccupied band. In contrast, the electrons can be scattered from the lower to upper bands for $\hbar/\tau > \Delta^{gap}$, as shown in Fig. S3B. Such a scattering changes the integrated Berry curvature over the Fermi sea, which suppresses the dissipationless intrinsic AHE (Fig. S3C). Meanwhile, the impurity scattering also induces both the extrinsic AHE caused by skew scatterings (inelastic scatterings) and the intrinsic contribution associated with the Fermi surface (i.e., dissipative AHE arising from incompletely filled bands (Fig. S3B)). As a result, the anomalous Lorenz number is suppressed for $\hbar/\tau > \Delta^{gap}$, leading to the breakdown of the Wiedemann–Franz law (*10*).

A similar behavior is expected to be observed for Weyl semimetals with time-reversal symmetry breaking, that is, magnetic Weyl semimetals (see Fig. S3, D to F). As shown in Fig. S3F, the energy dependence of intrinsic anomalous Hall conductivity is considered to become almost flat in the vicinity of the Weyl point (*25, 44*), indicating that the dissipationless intrinsic AHE is not affected by impurity scatterings as long as $\hbar/\tau$ is smaller than the cutoff energy, $\Delta^{cut}$, which is defined as the energy range of linear dispersion (Fig. S3D). In addition, it has been theoretically suggested that the extrinsic contribution and dissipative AHE coming from incompletely filled bands (Fig. S3E) identically vanish in the case of $\hbar/\tau < \Delta^{cut}$ (*26*). Thus, in the case of a small energy broadening, the anomalous Lorenz number maintains a constant value in accordance with the Wiedemann–Franz law.

### Section S5. Estimation of the scattering time from the Drude model

In conventional metals, the electric conductivity tensor is generally given by

$$\boldsymbol{\sigma} = \frac{e^2}{4\pi^3 \hbar} \int \frac{\boldsymbol{v_k v_k} \tau(\boldsymbol{k})}{v_k} dS_{\mathrm{F}},$$

where $S_{\mathrm{F}}$ is the area of the Fermi surface, *e* is the elementary charge, and $\tau(\boldsymbol{k})$ is the $\boldsymbol{k}$-dependent scattering time of charge carriers; moreover, $v_k = \frac{1}{\hbar}\frac{\partial \varepsilon_k}{\partial \boldsymbol{k}}$ is the velocity of the charge carriers, and $\boldsymbol{vv}$ represents the tensor, $\begin{pmatrix} v_x v_x & v_x v_y & v_x v_z \\ v_y v_x & v_y v_y & v_y v_z \\ v_z v_x & v_z v_y & v_z v_z \end{pmatrix}$. Assuming a single carrier and an isotropic system (namely, $\tau(\boldsymbol{k})$ = const. and $v_x = v_y = v_z$), we find that the electric conductivity is simply described by the single-carrier Drude model:

$$\sigma_{xx} = \frac{ne^2\tau}{m}.$$

Here, *n* is the density of states, and *m* is the effective mass of the charge carrier. Using $R_{\mathrm{H}} = 1/(ne) \sim 0.03 \times 10^{-2}$ cm$^3$/C (*15*) and $m = m_0$ ($m_0$ is the mass of free electron), we obtained the scattering rate, $\hbar/\tau^{Drude} = 0.35–0.88$ eV, from the single-carrier Drude model.

According to our first-principle calculations, however, Mn$_3$Sn has three bands with a non-negligible density of states near $E_{\mathrm{F}}$, one of which is a hole band with a density of states of $n_h = 1.38$ states/eV (band 50). The other two are electron bands with a density of states of $n_e^1 = 1.67$ states/eV (band 51) and $n_e^2 = 0.4$ states/eV (band 52) (see Fig. S4). Therefore, both the electron and hole bands should be considered in the calculation of the scattering time. Then, we assume the two-carrier Drude model, including the one electron (where the density of states is $n_e = n_e^1 +$ $n_e^2 = 2.07$ states/eV and the effective mass is $m_e = m_0$) and one hole band (where the density of states is $n_h = 1.38$ states/eV and the effective mass is $m_h = m_0$), for simplicity of discussion. In the two-carrier model, the electric conductivity is given by the sum of electronic ($\sigma_e$) and hole ($\sigma_h$) components,

$$\sigma_{xx} = \sigma_e + \sigma_h = \frac{n_e e^2 \tau_e^{Drude}}{m_e} + \frac{n_h e^2 \tau_h^{Drude}}{m_h},$$

where $\tau_e^{Drude}$ ($\tau_h^{Drude}$) is the scattering time of the electron (hole) band. In contrast, the Hall coefficient is given by

$$R_H = \frac{n_h \mu_h^2 - n_e \mu_e^2}{e(n_h \mu_h + n_e \mu_e)^2},$$

where $\mu_e = e\tau_e^{Drude}/m_e$ ($\mu_h = e\tau_h^{Drude}/m_h$) represents the mobility of electrons (holes). Using Eqs. S3 and S4, $\hbar/\tau_e^{Drude}$ and $\hbar/\tau_h^{Drude}$ are estimated to be 1.13–3.08 and 0.25–0.67 eV, respectively. Thus, the calculated $\hbar/\tau^{Drude}$, $\hbar/\tau_e^{Drude}$, and $\hbar/\tau_h^{Drude}$ are much larger than the $\Delta^{gap}$ of ~70 meV expected for conventional $3d$-ferromagnetic metals; hence, the model for ferromagnetic metals with a narrow gap induced by spin–orbit coupling (see Section S4) is likely to be incorrect for the present system.

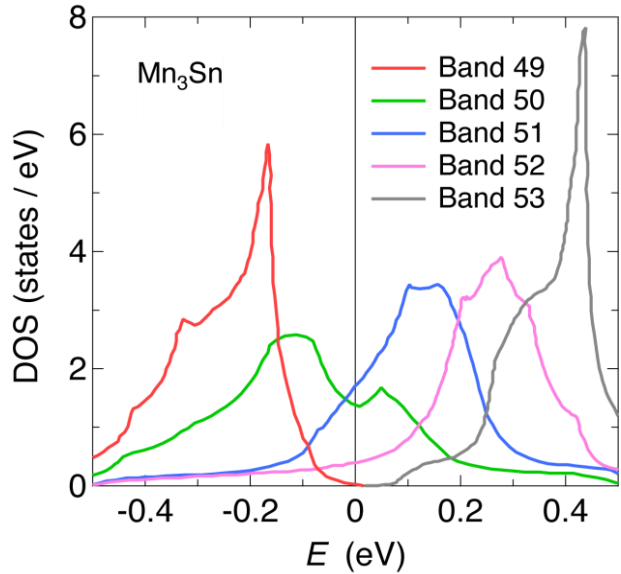

**Fig. S4. Energy dependence of the calculated density of states (DOS) near the Fermi energy ($E_F$).** For convenience, we set that $E_F$ of stoichiometric Mn$_3$Sn equals to $E = 0$ eV. In this case, $E_F$ of sample #1 (#2) becomes $E = 0.04$ (0.05) eV and the Weyl point is located at $E = 0.065$ eV. Two electron bands (blue and pink lines) and one hole band (green line) have a large density of states near the Weyl point.

**Section S6. Estimation of the scattering time of Weyl bands**

To estimate the value of $\tau^{Weyl}$, we focus on the Mn doping dependence of $L_{zx}^{AH}$ for Mn$_3$Sn. Recently, it has been suggested that the energy difference between Weyl points and $E_F$ of sample #1 (sample #2) is $\Delta E^{Weyl} \sim 25$ meV (~15 meV) (*16*) and the energy range for linear dispersion is $\Delta^{cut} \sim 40$ meV (*20, 23, 24*). For sample #1, the value of $L_{zx}^{AH}$ slightly deviates from $L_0$ above $T \sim 50$ K (Fig. 4A), which indicates that $\hbar/\tau^{Weyl}$ is larger than ~14.4 meV because the relation $\sqrt{(\hbar/\tau^{Weyl})^2 + (k_BT)^2} \geq \Delta^{cut} - \Delta E^{Weyl}$ holds at $T \geq 50$ K. In contrast, the value of $L_{zx}^{AH}$ for sample #2 retains the Wiedemann–Franz law at least up to $T = 150$ K (Fig. 4A), which means that $\hbar/\tau^{Weyl}$ is estimated to be less than ~21.4 meV by assuming that $\sqrt{(\hbar/\tau^{Weyl})^2 + (k_BT)^2} \leq \Delta^{cut} - \Delta E^{Weyl}$. On the basis of these two equations, $\hbar/\tau^{Weyl}$ can be roughly estimated to be ~14.4–21.4 meV. Surprisingly, this value is one or two orders of magnitudes lower than $\hbar/\tau^{Drude}$ (see Section S5), indicating that the scattering probability between the Weyl bands and metallic bands is significantly suppressed. Further theoretical studies are required to determine the coexistence of the Weyl bands and metallic bands.